\documentclass{article}
\usepackage{arxiv}
\usepackage[utf8]{inputenc} 
\usepackage[T1]{fontenc}    
\usepackage{url}            
\usepackage{booktabs}       
\usepackage{amsfonts}       
\usepackage{nicefrac}       
\usepackage{microtype}      
\usepackage{graphicx}
\usepackage{natbib}
\usepackage{doi}
\usepackage{lineno,hyperref,soul, subfig, tikz}

\title{Automatic Feature Detection in Lung Ultrasound Images using Wavelet and Radon Transforms}

\date{} 					

\author{ 
Maria Farahi \footnote{maria.farahi@sanameditech.com} \\ 
Sana Meditech S.L. \\
08014, Barcelona, Spain\\
Research Center in Biomedical Engineering (CREB)\\
Universitat Polit`ecnica deCatalunya\\
08034, Barcelona, Spain\\
	\And
Joan Aranda \\
Research Center in Biomedical Engineering (CREB)\\
Universitat Polit`ecnica de Catalunya\\
08034, Barcelona, Spain\\
     \And
Hessam Habibian \\
Sana Meditech S.L. \\
08014, Barcelona, Spain\\
     \And
Al\'{i}cia Casals \\
Research Center in Biomedical Engineering (CREB)\\
Universitat Polit`ecnica de Catalunya\\
08034, Barcelona, Spain\\
}
\hypersetup{
pdftitle={Automatic Feature Detection in Lung Ultrasound Images using Wavelet and Radon Transforms},
pdfsubject={eess.IV},
pdfauthor={Maria Farahi, Joan Aranda, Hessam Habibian, and Al\'{i}cia Casals},
pdfkeywords={Image Processing, Lung Ultrasound, LUS, Pattern Extraction, Pattern Localization, Radon Transform, Wavelet Analysis.}
}
\begin{document}
\maketitle

\begin{abstract}
	{\bf Objective:} Lung ultrasonography is a significant advance toward a harmless lung imagery system. This work has investigated the automatic localization of diagnostically significant features in lung ultrasound pictures which are Pleural line, A-lines, and B-lines. {\bf Study Design:}  Wavelet and Radon transforms have been utilized in order to denoise and highlight the presence of clinically significant patterns. The proposed framework is developed and validated using 3 different lung ultrasound image datasets. Two of them contain synthetic data and the other one is taken from the publicly available POCUS dataset. The efficiency of the proposed method is evaluated using 200 real images.  {\bf Results:} The obtained results prove that the comparison between localized patterns and the baselines yields a promising F2-score of 62\%, 86\%, and 100\% for B-lines, A-lines, and Pleural line, respectively. {\bf Conclusion:} Finally, the high F-scores attained show that the developed technique is an effective way to automatically extract lung patterns from ultrasound images.
\end{abstract}


\section{Introduction}
are different types of Lung Diseases (LD) including pneumonia, pneumothorax, SARS, COVID-19, etc. \cite{alvarado2016metabolic}. LD pathologies could worsen and cause more severe problems if left without treatment. For example, untreated pneumonia would give rise to a blood infection or even lung abscesses \cite{yoon2013tension}. Similarly, unhealed pneumothorax or COVID-19 might lead to death \cite{zhu2020novel}.

At present, there are several ways to perform lung tests including computed tomography (CT) \cite{yau2021point}, chest X-ray \cite{buonsenso2020novel},  magnetic resonance imaging (MRI) \cite{sodhi2021lung}, and ultrasound (US) \cite{lichtenstein2014lung}. CT imaging is a gold standard for lung pathologies detection since it generates high-resolution images. However, it is expensive and delivers a high dose of radiation to the patient \cite{yau2021point}. X-ray is widely available and cost-effective although it can not assure the detection of all lung pathologies and still produces ionizing radiation \cite{buonsenso2020novel}.  Advances in MRI have made it a feasible option for lung evaluation. MRI yields very high-resolution images without ionizing radiation. However, it is expensive and with limited availability \cite{sodhi2021lung}.

Instead, US avoids radiation and is a cheaper modality \cite{lichtenstein2014lung}. It does not require special shielded and equipped rooms like  MRI or CT \cite{shams2021lung}. It also can be effectively utilized by the bedside, thus avoiding the need to move the patient and consequently, saving time and cost.\cite{buonsenso2020novel}. However, US images have not been used widely for classification approaches because of their low-quality 
\cite{liu2019deep}. In spite of this limitation, this work will prove that US is reliable enough to make a first lung disease diagnosis.

The emergence of the COVID-19 pandemic has pushed the use of ultrasonography devices \cite{soldati2020there}. This shift in focus is primarily due to the time-consuming nature of CT and MRI scans. As a consequence, healthcare professionals have increasingly relied on ultrasound images to feed their decision-making processes, aiming to minimize the necessity for unnecessary hospitalizations \cite{smargiassi2021lung}. They follow the Bedside Lung Ultrasound in Emergency (BLUE) protocol  \cite{lichtenstein2014lung,bekgoz2019blue} to make their decisions. The BLUE protocol is a systematic approach used by clinicians to evaluate and diagnose various lung pathologies, such as pneumothorax, pneumonia, pleural effusion, etc. by examining specific lung zones and patterns using lung ultrasound imaging. The LUS patterns include the Pleural line, A-lines, and B-lines. 

In an adult LUS image, the Pleural line is formed by the interface between the air-filled lung tissue and the pleural membrane. The ultrasound waves encounter this boundary and undergo a strong reflection, resulting in a bright horizontal line on the image. The pleural line is an essential landmark in lung ultrasound as it helps identify the lung boundaries and serves as a reference point for further analysis.

A-lines are created due to the multiple reflections of ultrasound waves between the pleural line and the ultrasound probe. As the sound waves bounce back and forth, they create parallel lines that appear equidistant from each other. A-lines are typically observed in normal lung tissue and indicate a healthy air-filled lung with no significant pathology.

\begin{figure*}[!t]
\centerline{
\subfloat[]{\includegraphics[width=0.24\columnwidth]{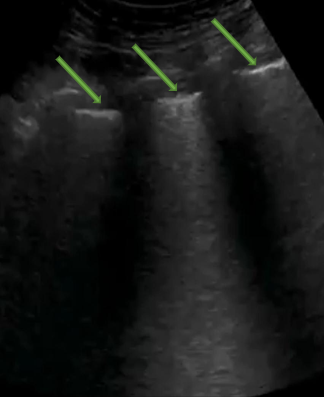}
\label{fig:normalLung}}

\hfil
\subfloat[]{\includegraphics[width=0.24\columnwidth]{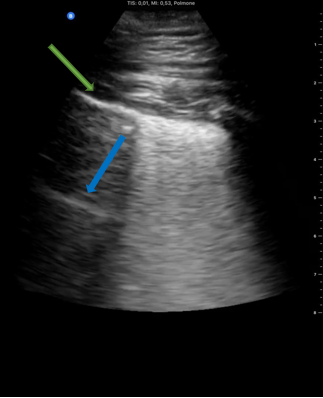}
\label{fig:normalLung1}}
\hfil
\subfloat[]{\includegraphics[width=0.24\columnwidth]{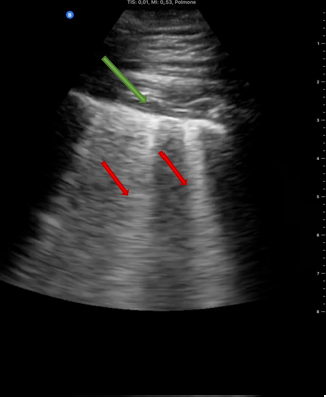}
\label{fig:covidLung}}
\hfil
\subfloat[]{\includegraphics[width=0.24\columnwidth]{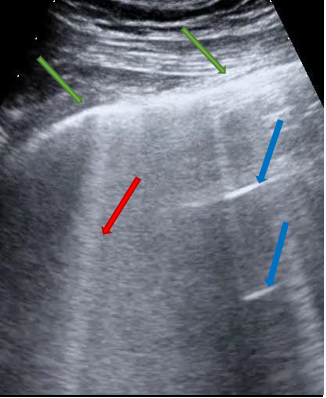}
\label{fig:pneumoniaLung}}
\hfil}
\caption{Adult's lung ultrasound images and their patterns indicating the Pleural line (green arrows), A-lines (blue arrows), and B-lines (red arrows). LUS imafes including (a) Only Pleural line, (b) Pleural line and A-Line, (c) Pleural line and B-Line, (d) Pleural line, A-Lines, and B-Lines.}
\label{fig:LUSPatterns}
\end{figure*}

B-lines occur when ultrasound waves encounter a significant difference in tissue density, such as the presence of fluid in the lung interstitium or alveoli. The sound waves penetrate the fluid-filled spaces and are strongly reflected, resulting in vertical bright lines starting from the pleural line. They are commonly associated with various pathological conditions, such as pulmonary edema, interstitial lung diseases, or pneumonia. The number and distribution of B-lines can provide important diagnostic information about the underlying lung pathology. 
\cite{lichtenstein2014lung}. Fig. \ref{fig:LUSPatterns} shows these LUS features in a real image taken from an adult's lung using a convex probe. Fig. \ref{fig:normalLung} is a normal lung showing pleural line withous any A or B lines. Fig \ref{fig:normalLung1} is a normal lung showing one A-line. Fig \ref{fig:covidLung} belong to a patient suffering from COVID-19 showing multiple B-Lines. Finally Fig \ref{fig:pneumoniaLung} is from a pneumonia patient with one -Line and two -Lines. Table \ref{tab:PathologiesCharacteristics} illustrates how the mentioned features are useful in lung ultrasound disease diagnosis.

\begin{table}[!t]
    \caption{A summary of LUS features in different lung pathologies}
    \begin{tabular}{p{0.4\columnwidth}p{0.07\columnwidth}p{0.07\columnwidth}p{0.07\columnwidth}p{0.07\columnwidth}}
 \hline
 
    \textbf{\small{\rotatebox[origin=c]{0}{Lung Pathology}}}  &  \textbf{\small{\rotatebox[origin=c]{-90} {Ribs' Shadow}}}& \textbf{\small{\rotatebox[origin=c]{-90}{Pleural Line}}} &\textbf{\small{\rotatebox[origin=c]{-90}{A-Lines}}} & \textbf{\small{\rotatebox[origin=c]{-90} {B-Lines}}} \\
    \hline

       Normal Lung  & \checkmark &\checkmark &  * & * \\
       \hline
    Pneumothorax  &  \checkmark & \checkmark  & \checkmark& \texttimes \\
  
      \hline
    COVID-19  & \checkmark & \checkmark &  * &$\geq 2 $  \\
   
      \hline
   Broncho-Pneumonia & \checkmark & \checkmark & \checkmark & \checkmark \\
  
  \hline
   Pneumonia Consolidation& \checkmark & \texttimes & \texttimes& \textminus \\
    \hline
    Pleural Effusion & \checkmark & \texttimes & \texttimes & \texttimes  \\
    \hline

    \end{tabular}\\
    \small {*May be seen.}
    \label{tab:PathologiesCharacteristics}
\end{table}
Nowadays, Deep Learning (DL) methods are commonly used for image segmentation and diagnosis \cite{bandyk2021mri}. In DL, features are automatically extracted by a network after a training process. Deep learning-based systems come out with good performance provided that the amount of available data is large enough \cite{bandyk2021mri}. However, they are black boxes, which often lead to unexplainable procedures. They also require special and expensive GPU systems for training. Instead, standard computer vision algorithms are preferred when data is scarce due to their ability to achieve reliable results with smaller amounts of data. These algorithms typically follow logical frameworks that are explainable and understandable, making them easier to interpret and analyze. This makes them practical and widely applicable in various computer vision tasks. Nevertheless, exploring the best way for their formulation is not always easy and fast, or even feasible \cite{khan2021machine}. Taking into consideration the lack of a considerable amount of LUS images, in the current study, we have explored the feasibility of relying on computer vision.



During the past decade, especially after the appearance of COVID-19, some efforts were dedicated to discover a suitable way for LUS patterns extraction. For instance, Barrientos \emph{et al.} \cite{barrientos2016filtering} utilized US image data from children under 5 suffering from pneumonia to detect the Pleural line. The image was locally thresholded using a vertical gradient. They compared their final results with manually segmented images, reporting a mean quadratic error of 11.17 pixels. Karakus \emph{et al.} \cite{karakucs2020detection} combined a sparsity-enforcing and Cauchy-based penalty process to quantify B-lines in LUS images of patients with COVID-19. They asserted that their algorithm correctly detects B-lines in 87\% of non-COVID-19 patients.  Unfortunately, they did not reveal any results about the testing of their algorithm on patients suffering from COVID-19. Susanti \emph{et al.} \cite{susanti2021image} utilized a top-hat morphological grayscale filtering method with a texture structure element to extract the Pleural line and A-lines. They also did not report any sensitivity or specificity of their results.

Anantrasirchai \emph{et al.} \cite{anantrasirichai2017line} proposed a method for line detection in speckle images. They used the Radon transform and the L1 regularization method to extract lines from 50 simulated noisy LUS images. They reported the simulated B-lines detection performance of 54\%, 40\%, and 33\% for F0.5, F1, and F2 scores, respectively. Table~ \ref{tab:SummeryStateOfTheArt} summarizes the mentioned works on LUS pattern extraction.

\begin{table}
    \setlength{\tabcolsep}{4pt}
    \centering
    \scriptsize
    \caption{Summary of the works mentioned in the state of the art}
    \begin{tabular}{|p{0.3cm}|p{2.5cm}|p{2.5cm}|p{2.5cm}|}
    
    \hline
          Ref. & Goal &  Data & Result  \\
         \hline
          \cite{li2010fetal}& Lung area segmentation& LUS Images  from normal pregnant women.& Shown by images. \\
          \hline
          \cite{barrientos2016filtering}& Skin elimination in LUS images  & 23 LUS images from children under 5. & Mean quadratic error = 11.17 pixels \\
          \hline
          \cite{karakucs2020detection}& Line artifacts quantification in LUS & LUS images from 9 patients with COVID-19  &  87\% correctly detected lines in non-COVID samples. \\
          \hline
          \cite{susanti2021image}&  Pleural line and A-line detection & 150 LUS frames of public data &  Shown by images. \\
           \hline
          \cite{anantrasirichai2017line}& Line restoration in speckle images  & 50 simulated LUS images &  F0.5=54\%,  F1=40\%, and F2=33\%  \\
          \hline
    \end{tabular}
    
    \label{tab:SummeryStateOfTheArt}
\end{table}

 With the aim to improve these still poor results on segmentation, we have developed an algorithm based on the use of Wavelet to denoise and Radon transforms to extract lung ultrasound features including Pleural line, A-Lines, and B-Lines. In section II we illustrate the materials and data conditioning. Our proposed method for data analysis is explained in section III. Results are shown in section IV and the results and method performance are discussed in section V.

\section{Materials and Data Conditioning}
\begin{figure}
    \centering
    \includegraphics[ scale=0.5]{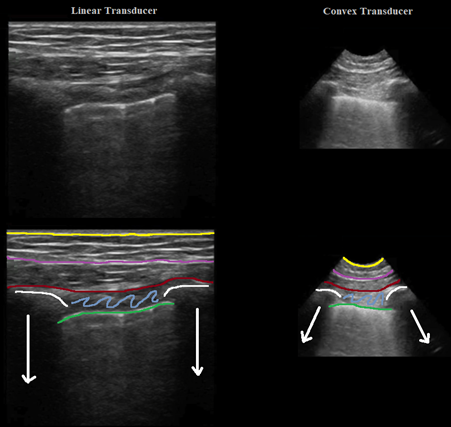}
    \caption{\small LUS layers. The first and second columns show images from a linear and a convex transducer respectively. The skin layer (yellow), the subcutaneous layer (purple), the muscular layer (red), the cortex of the ribs (white curve) with shadowing  (white arrow), the intercostal region between the ribs (blue), and the pleural line (green) are all depicted in the second row}
    \label{fig:LUS-Layers}
\end{figure}

In a LUS image, all clinically important signs arise below the Pleural line \cite{lichtenstein2014lung}. When the Pleural line is detected, the area above can be discarded cropping the image since it just represents the skin and muscles. Indeed, significant information on the A and B lines appears below the Pleural line. Fig. \ref{fig:LUS-Layers} shows the layers of a LUS image. Observing the following clinical features is beneficial in order to extract relevant patterns \cite{lichtenstein2014lung}.
\begin{itemize}
    \item From the top of the image, a LUS image starts with information about skin and muscles.
    \item The Pleural line appears after ribs.
    \item The Pleural line is horizontal with an angle between -15 to +15 degrees.
    \item A-lines are the reflection of a Pleural line and appear at the same distance from each other
    \item B-lines are vertical and start from Pleural line and advance down through the image to the bottom
    
\end{itemize}

 Knowing the characteristics of these patterns, we use three different data-sets to develop the framework.

\subsection{Dataset}
To design the denoising procedure, we created a dataset consisting of 1950 simulated ultrasound images. These images were generated using the Matlab Ultra Sound Toolbox (MUST). We modeled a convex transducer with 76 elements. The dataset was designed to mimic real-world ultrasound data and encompassed a variety of parameters, including the number of A-lines (from 0 to 3), B-lines (from 0 to 5), and pleural lines (one/multiple). The simulated data covered a range of US frequencies (from 1MHz to 5MHz) and depths (from 7cm to 11cm), representing different imaging scenarios. By incorporating various frequencies, we aimed to capture the diversity of ultrasound signals encountered in clinical practice. Additionally, the different depths simulated in the dataset allowed us to evaluate the performance of the denoising methodology across different imaging depths. The total 1950 images used contain 3300 Pleural lines, 1575 A-Lines, and 4500 B-Lines.

\begin{figure}
    \centering
    \includegraphics[scale=0.3]{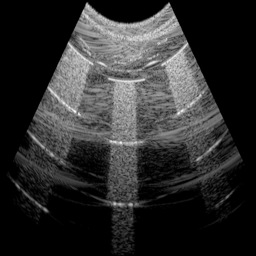}
    \includegraphics[scale=0.3]{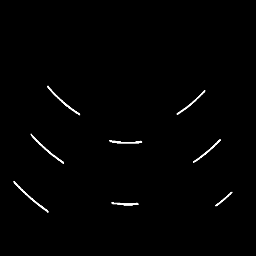}
    \includegraphics[scale=0.3]{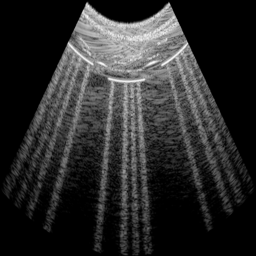}
    \includegraphics[scale=0.3]{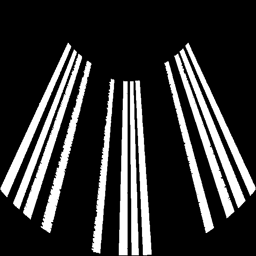}
    
    \caption{Simulated samples taken from  \cite{zhao2022covid}.From left to right: simulated image with A-lines, related mask to A-lines, simulated image with B-lines, and related mask to B-lines.}
    \label{fig:simulatedSamples}
\end{figure}

We used the simulated dataset published in \cite{zhao2022covid} to develop the features detection procedure. In this dataset, 30,000 lung phantoms were simulated with MATLAB using Field II simulation software package. Simulations were based on lung ultrasound images with A-line, B-line, and consolidation features. We used data related to A-line (10,000 images) and B-line (10,000 images) to implement our algorithm. Fig \ref{fig:simulatedSamples} shows samples and their related ground truth of this data set. The left image contains the simulated image while the right image depicts their ground-truth mask.

To validate the proposed framework, we utilized a publicly available dataset, POCUS, consisting of different lung pathologies, like pneumonia, COVID-19, and healthy \cite{born2021accelerating}. It includes 183 video clips acquired by different types of transducers. We also utilized published data in \cite{ketelaars2018ultrasound} that contains 32 convex video clips recorded from patients suffering from pneumothorax. From these data, we randomly chose 200 frames to test our algorithm. Table \ref{tab: studyPopulationCriteria} shows the distribution of the used data per disease category. Since the used dataset is provided by different hospitals that apply various convex transducers in different frequencies, we manually selected the ultrasound area of the images to normalize.


\begin{table}
\setlength{\tabcolsep}{6pt}
\scriptsize
\caption{The distribution of public data used for the study}
\label{tab: studyPopulationCriteria}
\centering
\begin{tabular}{|p{1.3cm}|p{1cm}|p{1cm}|p{2cm}|p{0.5cm}|}
\cline{2-4}
\multicolumn{1}{p{1.3cm}}{}&\multicolumn{2}{|p{2cm}}{ No. of Data} &  \multicolumn{1}{|p{2cm}|}{ No. of Used Frames} &\multicolumn{1}{p{0.5cm}}{} \\
\hline
Group Name& Videos & Frames & Test our work & Ref.\\
\hline
{Normal Lung} & 28 & 3834  & 50&\cite{born2021accelerating}\\
\hline
{COVID-19} & 54 & 10083  &50 &\cite{born2021accelerating}\\
\hline
{Pneumonia} & 38 & 4254  & 50&\cite{born2021accelerating} \\
\hline
{Pneumothorax} & 32 & 14400  &50 &\cite{ketelaars2018ultrasound}\\
\hline
{Sum} &  152 & 32571  & 200 &\multicolumn{1}{|c}{}\\
\cline{1-4}
\end{tabular}
\end{table}

\subsection{Pre-Processing and denoising}

Ultrasound images are susceptible to heavy corruption caused by various sources of noise, including patient movements, sensor movements, and reflections from the skin and muscles \cite{piscaglia2020benefits}. These noises can manifest as Salt and Pepper (impulse or spike), Poisson, Gaussian, and Speckle \cite{gupta2018performance}. In the field of ultrasound imaging, wavelet-based denoising methods are commonly employed \cite{georgieva2021medical}. For this, a comparative review performed by Georgieva et al. \cite{georgieva2020comparative} demonstrated the significant impact of using wavelet for noise reduction in ultrasound images. Additionally, in our previous study \cite{farahi2022beat}, we implemented a comprehensive methodology to identify the optimal wavelet family for denoising ultrasound signals obtained from a portable Doppler device. 

In this study, we utilize a methodology similar to that employed in our previous work to identify the optimal mother wavelet family for denoising lung ultrasound images. Our investigation involves assessing the Signal-to-Noise Ratio (SNR) of denoised images obtained using various mother wavelets, different levels of decomposition, and threshold values.

To explore the effectiveness of different mother wavelets, we consider a total of 61 options, including Haar, Daubechies (db1, db2, ..., db20), Symlets (sym1, sym2, ..., sym20), Coiflets (coeif1, coeif2, ..., coeif5), and Biorthogonal (bior1.1, bior1.3, ..., bior6.8). Furthermore, we evaluate the impact of different levels of decomposition, ranging from level 2 to level 5. Additionally, we examine a wide range of threshold values, spanning from 0 to 101, to determine their influence on the denoising process. By systematically investigating these parameters, we aim to identify the combination that reaches the highest SNR after denoising.

\begin{figure}
    \centering
    \subfloat[]{\includegraphics[scale=0.5]{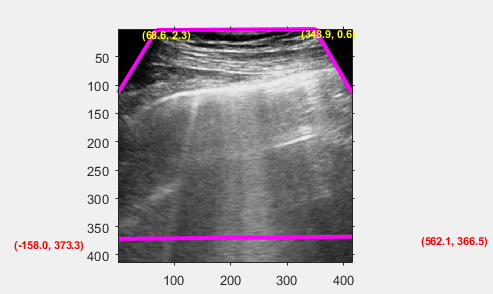}
    \label{fig:fig4a}}
    \subfloat[]{\includegraphics[scale=0.5]{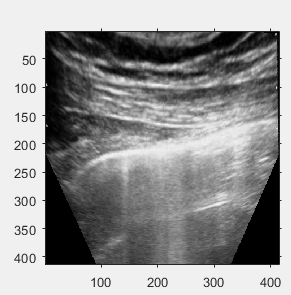}
     \label{fig:fig4b}}
    \caption{The result of transforming the convex image to cartesian using the affine transformation. (a) Convex denoised image and the deployed trapezoid on it. (b) Transformed trapezoid to rectangular.}
    \label{fig:convex2linear}
\end{figure}

In order to facilitate further analysis and processing, we perform a transformation on the ultrasound images obtained from convex transducers. Since the original images are in polar format, we employ a trapezium to delineate the LUS area and subsequently apply an affine transformation to convert the images into a rectangular format. This transformation enables us to work with the LUS data more conveniently in Cartesian coordinates, allowing for an easier interpretation and analysis. (Fig. \ref{fig:fig4a}, \ref{fig:fig4b}).

\section{Data Analysis}
Data analysis of LUS images involves processing and interpreting ultrasound area to extract meaningful information about the lungs. It mainly includes tasks such as Pleural line, A and B lines extraction which are explained below. 
\subsection{Pleural-line extraction}
Pleural  line detection and localization
in lung ultrasound images is a challenging task due to factors such as noise, low image quality, and variability in the appearance of the pleural line across different patients. Once discarded both, the use of gradients to segment the detected different tissue structures and the use of DL due to the lack of enough data, it's essential to use advanced image processing and computer vision techniques, as well as a reasonable dataset for training and validation to achieve good performance.
\begin{figure*}[!t]
    \centering
    {\includegraphics[scale=0.55]{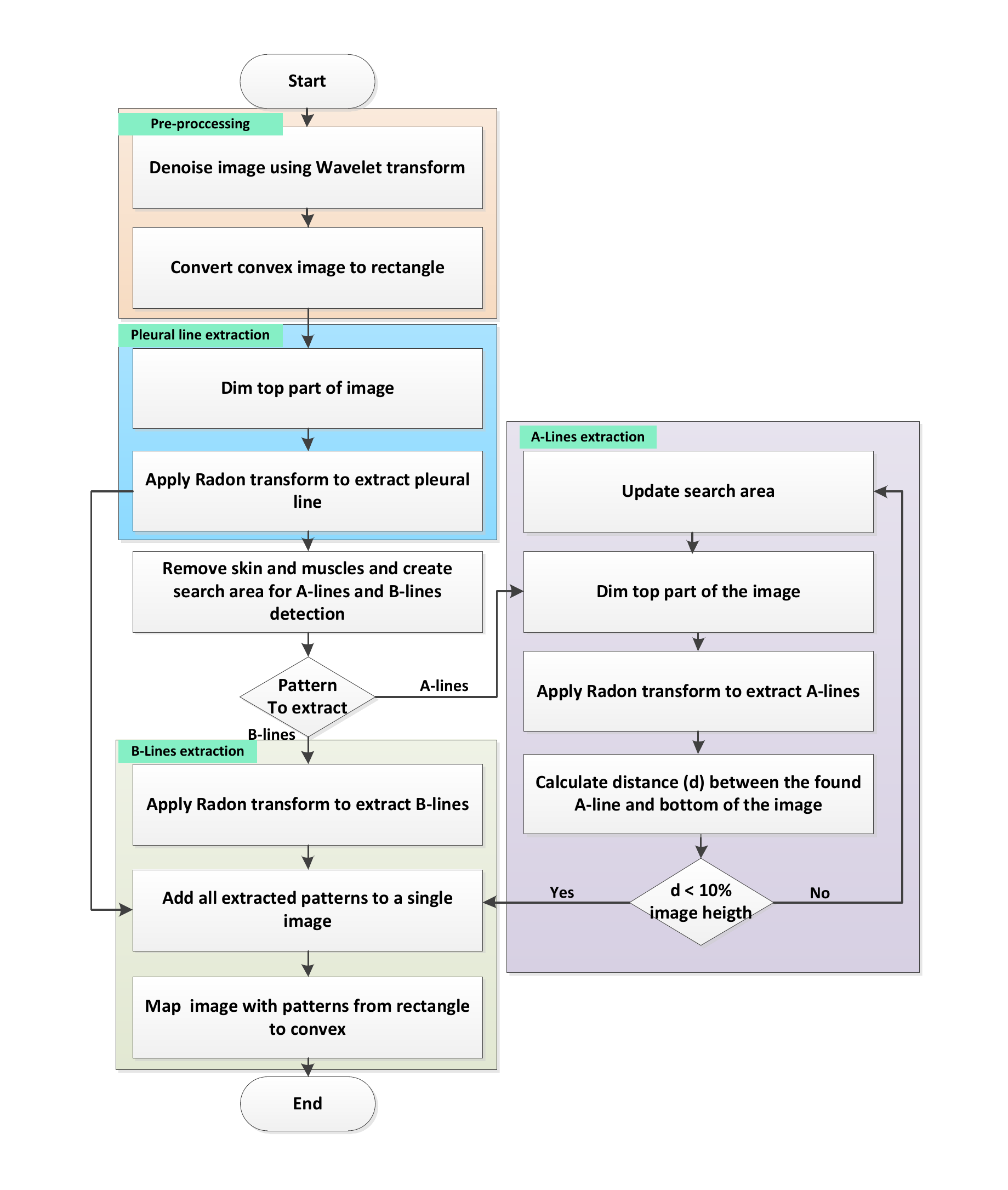}}
    \caption{The proposed framework to extract lung features including Pleural line, A-lines and B-lines.}
    \label{fig: PatternExtractionSchema}
\end{figure*}
We employ a series of steps to extract the Pleural line. Firstly, we apply a blurring technique to reduce the influence of skin and muscle in the top part of the image. This helps to enhance the visibility of the Pleural line, which typically exhibits a slope within the range of [-15, 15]+$\frac{\pi}{2}$ degrees \cite{lichtenstein2014lung}. Next, we apply the Radon transform values within the range of [-20, 20]+$\frac{\pi}{2}$ degrees, as this range considers the potential orientations of the Pleural line. By applying the Radon transform, the image is transformed from the spatial domain to the Radon domain, where straight lines appear as distinct peaks. This enables efficient detection and localization of the Pleural line. By identifying the local maximum intensity in the Radon domain, we are able to locate the Pleural line. It is worth to mention that the Pleural line is characterized by its length and brightness, making it distinguishable from other structures in the image. 

Once the Pleural line is detected, we proceed to ignore the skin and muscle regions from the image. This allows us to focus our subsequent analysis on the bottom region of the Pleural line, where important features related to lung abnormalities are more likely to be present. For a visual representation of the applied steps in Pleural line extraction, refer to Fig. \ref{fig: PatternExtractionSchema}, which provides a detailed overview of the process.




\subsection{A-lines and B-lines extraction}
To extract A-lines, which are reflections of the Pleural lines, we utilize the Radon transform with a narrow angle range between $(\theta _ {PL} -5) $ and $(\theta _ {PL} +5) $ degrees. Here, $\theta _ {PL} $ represents the angle of the extracted Pleural line. Since A-lines are short straight lines, the procedure involves applying the Radon transform to each A-line and discarding the top part of the line. This process is repeated iteratively until the bottom of the image is reached, enabling the extraction of A-lines throughout the lung region of interest. Finally, we check the distance between the extracted A-lines. It should be equal to the distance between the pleural line and the skin. A-lines that do not have a similar distance are discarded. 

Similarly, for B-line extraction, we employ the Radon transform on a skin-removed image because B-lines are also straight bright lines. In this case, the transform is applied once, with a vertical angle range between [-5, 5] degrees. This allows us to capture the characteristic patterns associated with B-lines in the image. In the end, B-lines that are not expanded from the Pleural line to the bottom of the LUS area are discarded.

To provide an overview of the LUS pattern extraction process, please refer to Fig. \ref{fig: PatternExtractionSchema}, which illustrates the schema of the procedure.

\subsection{Evaluation Metrics}
\begin{figure}
    \centering
    {\includegraphics[scale=0.35]{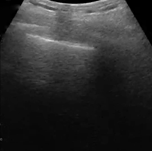}}
    {\includegraphics[scale=0.35]{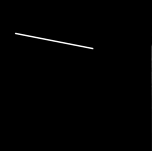}}
    {\includegraphics[scale=0.35]{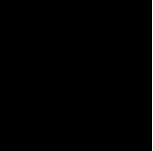}}
    {\includegraphics[scale=0.35]{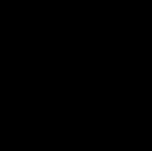}} \\

    {\includegraphics[scale=0.35]{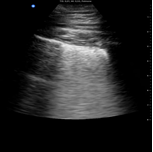}}
    {\includegraphics[scale=0.35]{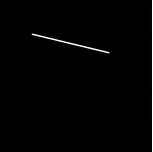}}
    {\includegraphics[scale=0.35]{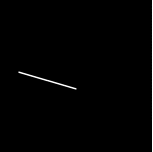}}
    {\includegraphics[scale=0.35]{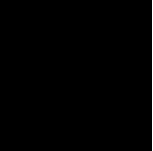}} \\
    
    {\includegraphics[scale=0.35]{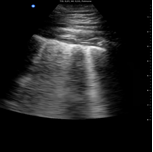}}
    {\includegraphics[scale=0.35]{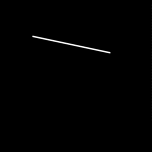}}
    {\includegraphics[scale=0.35]{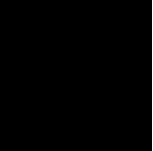}}
    {\includegraphics[scale=0.35]{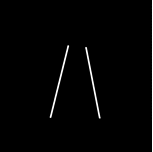}} \\
    
    {\includegraphics[scale=0.35]{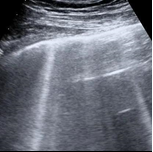}}
    {\includegraphics[scale=0.35]{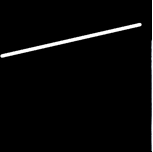}}
    {\includegraphics[scale=0.35]{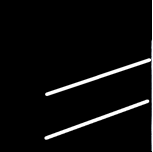}}
    {\includegraphics[scale=0.35]{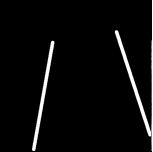}}
    
    \caption{Different LUS samples and their manually extracted masks as ground truth. The first column shows LUS images. The second column illustrates the Pleural line, while the third illustrates A-lines and the last column depicts B-lines.}
    \label{fig:masks}
\end{figure}

To examine the performance of pattern detection and localization, we structured a ground truth by manually extracting the existing patterns and creating masks for the Pleural line, A and B-lines (see Fig. \ref{fig:masks}). The manual extraction of the masks was performed by an expert in the field and verified by a physician to ensure their accuracy. By comparing the algorithm's results with the ground truth, we were able to determine the precision and recall of the algorithm's localization performance. This analysis helps us assess how accurately the algorithm is able to identify and locate the patterns of interest.

In addition to precision and recall, we calculated the top three common F-scores (F0.5, F1, and F2) \cite{jiang2020statistical}, which provide a comprehensive evaluation of the algorithm's performance by considering both precision and recall simultaneously. These metrics allow us to assess the algorithm's effectiveness in detecting patterns in lung ultrasound images and provide insights into its potential clinical utility.


\section{Results}

\begin{figure}
    \centering
{\includegraphics[scale=0.25]{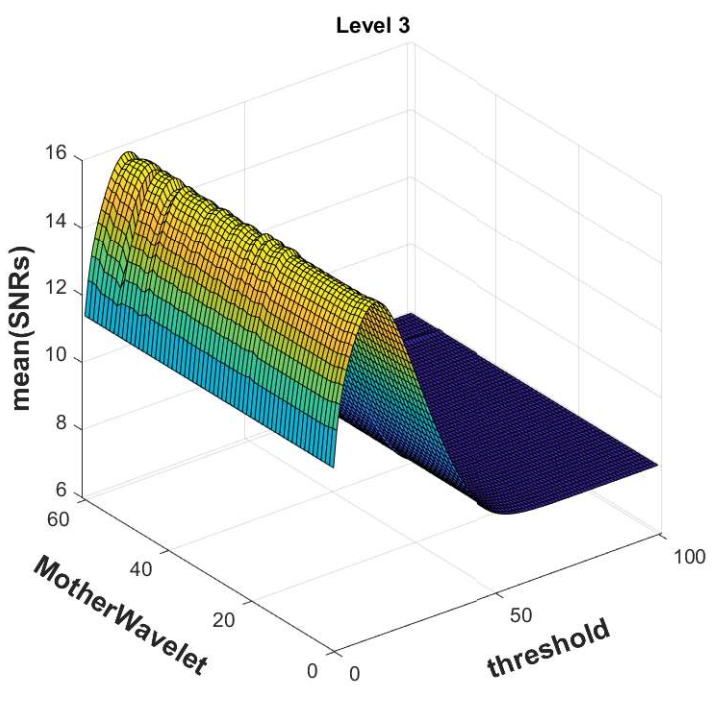}}
{\includegraphics[scale=0.25]{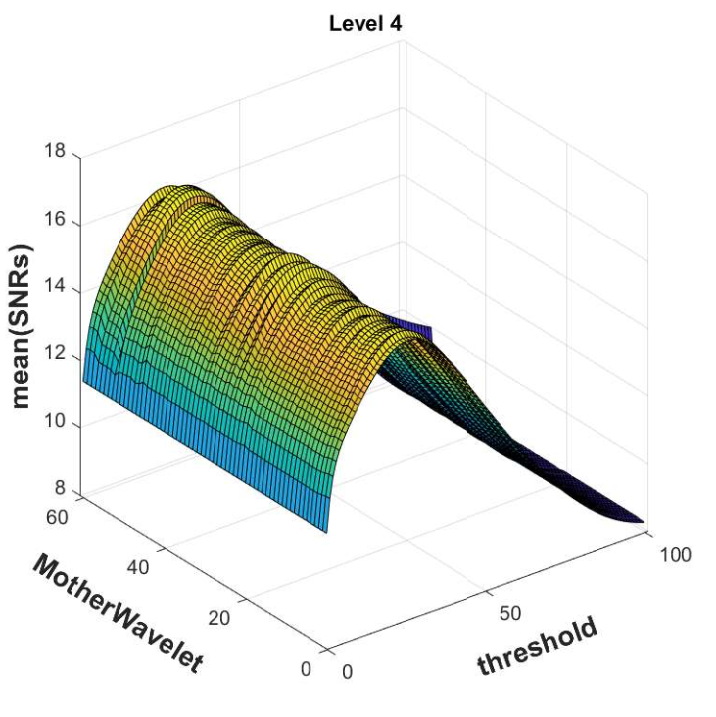}}
{\includegraphics[scale=0.25]{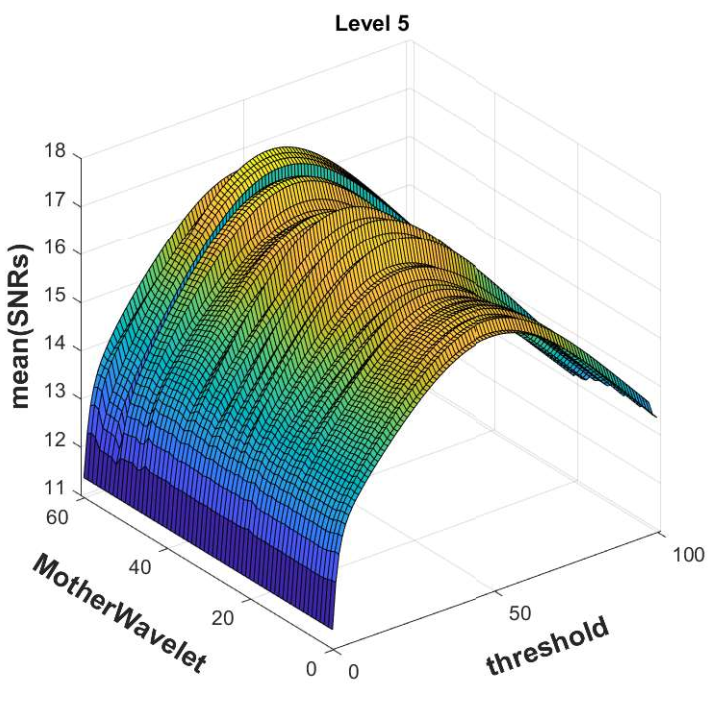}}
{\includegraphics[scale=0.25]{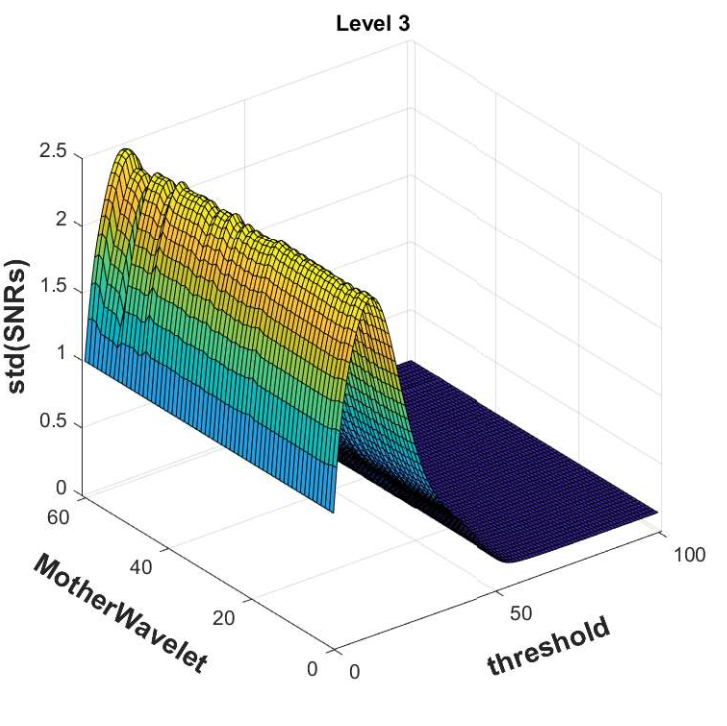}}
{\includegraphics[scale=0.25]{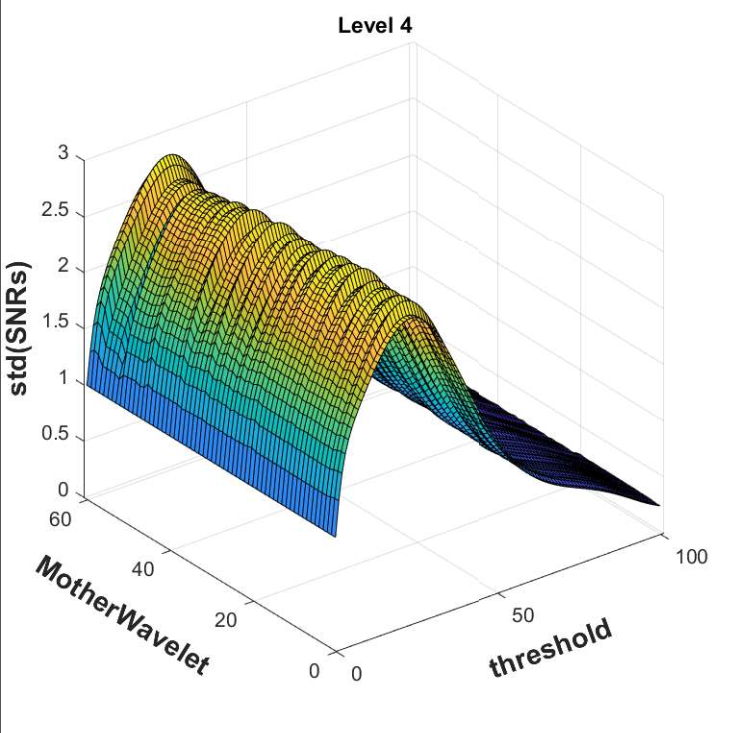}}
{\includegraphics[scale=0.25]{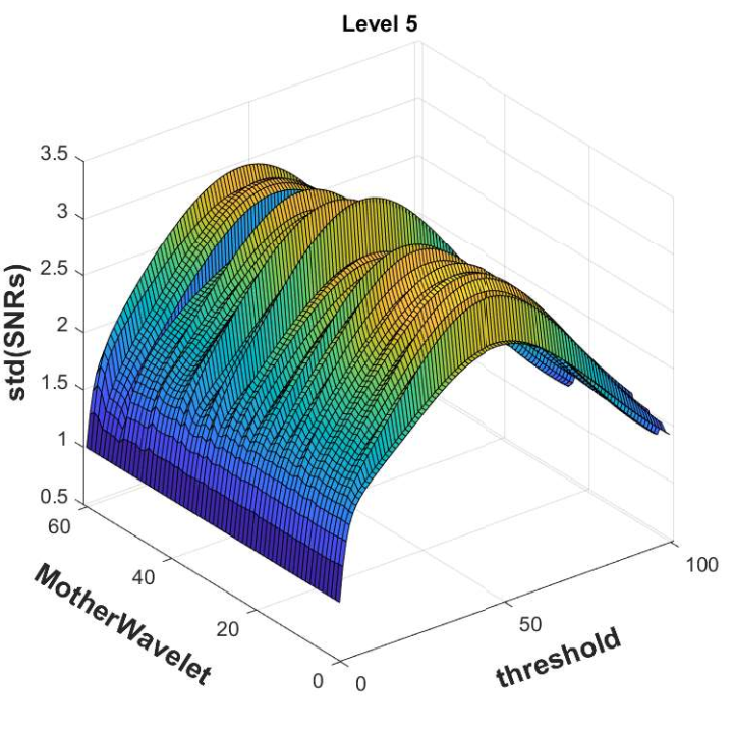}}
     \caption{Average and standard deviation of SNRs obtained after denoising using mother wavelets. The first row shows the average of SNRs wile the standard deviation of the SNRs is depicted in the second row. Level of decomposition 2, 4, and 5 are shown in the first, second, and third column, respectively. In each graph, xlabel depicts the threshold, and ylabel is the chosen mother wavelet. }
    \label{fig: stdAvgHeatMap}
\end{figure}

 In this paper, we denoised LUS images using different mother wavelets and analyzed their performance by calculating the SNR parameter. A higher value of SNR gives information about which type of mother wavelet is the mostc suitable for noise removal in LUS images.  Fig. \ref{fig: stdAvgHeatMap} provides information about the average and standard deviation of SNR on the different 1950 simulated images. As shown in Fig. \ref{fig: stdAvgHeatMap}, the average value of the SNRs is higher in level 5. Also, the 5th level is less sensitive to the threshold selection. 
 
\begin{figure}
    \centering
     {\includegraphics[width=0.32\columnwidth]{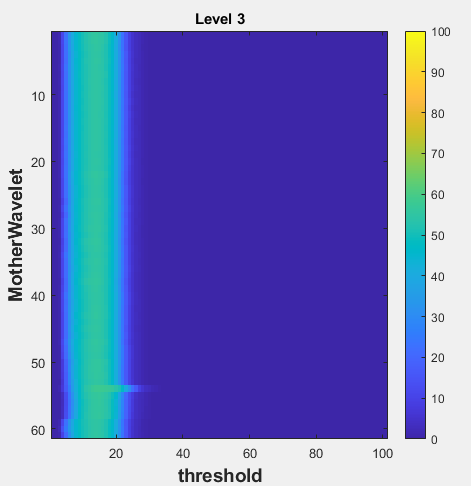}}  {\includegraphics[width=0.32\columnwidth]{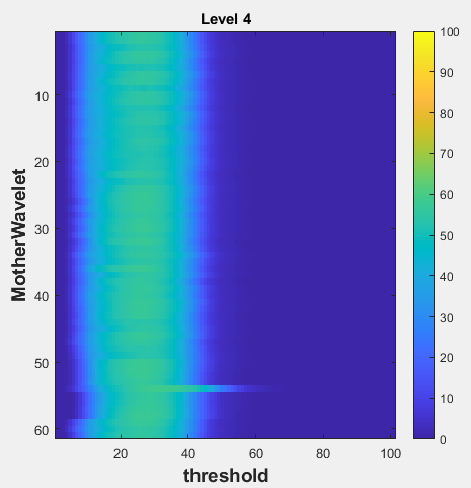}}   {\includegraphics[width=0.32\columnwidth]{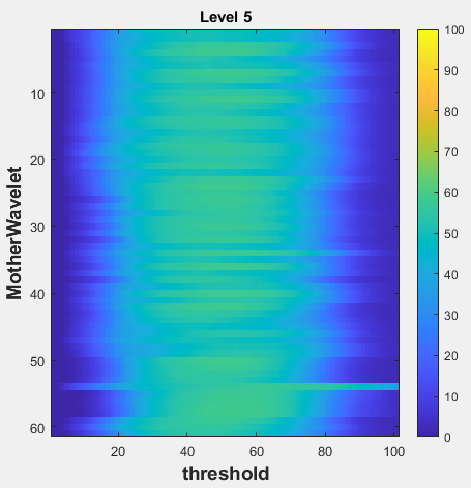}}
     \caption{Beter visualization of the number of images that has an 'SNR' more than the maximum of the 'average of SNRs'. Level 5 of decomposition is less sensitive to threshold selection than others.}
    \label{fig: SNRHeatMap}
\end{figure}

\begin{figure*}[!b]
    \centering    {\includegraphics[scale=0.3]{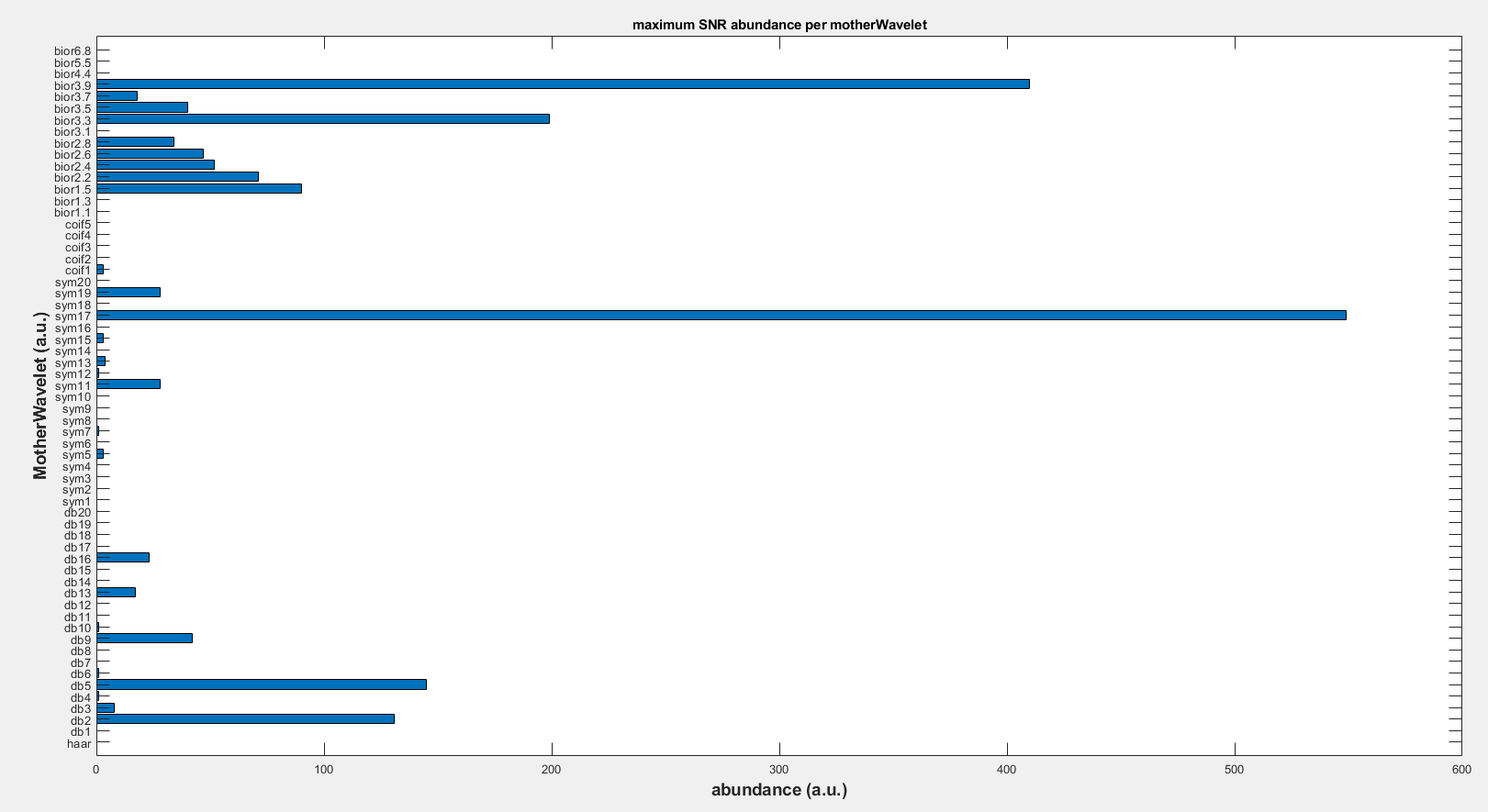}}
     \caption{Abundance of the occurrence of the maximum SNR in different mother wavelets. Most of the images yield their maximum SNR when 'Sym 17' is used.}
    \label{fig: MW61abundance}
\end{figure*}

 To have a better visualization, for each combination, the number of images that have an SNR higher than the maximum of the average of SNRs is calculated and shown in a heatmap graph (see Fig.\ref{fig: SNRHeatMap}). This figure explains that in level 5, the SNR is not highly sensitive to the selection of the threshold value, especially for values between 40 and 60. Therefore, in level 5 using a threshold of 50, we can choose the mother wavelet that more images show their maximum SNR with it. Based on the graph shown in Fig. \ref{fig: MW61abundance}, the 'sym 17' shows the maximum. So we chose the 'sym 17' with a threshold of 50 in the 5th level of decomposition to denoise the LUS images. 
 
  
\begin{figure}
     \centering
    \subfloat[]{ \includegraphics[scale=0.3]{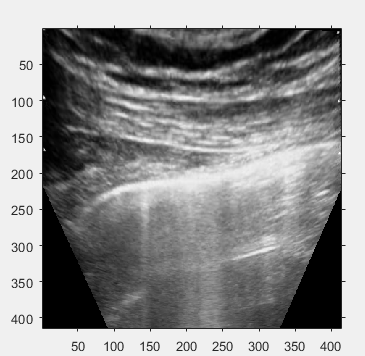}
    \label{fig:PLextractionResults1}}
     \subfloat[]{ \includegraphics[scale=0.28]{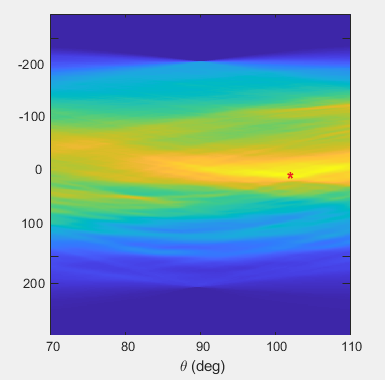}
     \label{fig:PLextractionResults2}}
     \subfloat[]{  \includegraphics[scale=0.3]{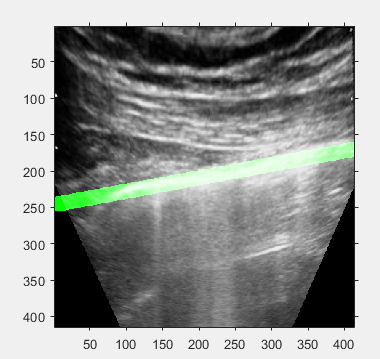}
     \label{fig:PLextractionResults3}}
     \subfloat[]{   \includegraphics[scale=0.3]{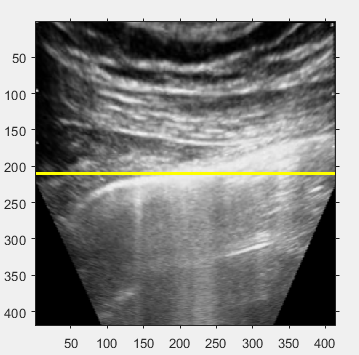}
     \label{fig:PLextractionResults4}}
        
     \caption{An example of pleural line extraction procedure. (a) Affined image to rectangular. (b) The rectangular image in Radon space. The maximum is shown by a red star. (c) extracted pleural line. (d) Split image using the obtained pleural line to define the lung area. The area below the yellow line is defined as the lung and is used for further steps.}
     \label{fig:PLextractionResults}
 \end{figure}

 Afterward, using local maxima in Radon space, we obtained the Pleural line position. Fig.\ref{fig:PLextractionResults} shows an example of pleural line extraction. In this figure, an affine transformation of an image from trapezium to rectangular is shown (Fig.\ref{fig:PLextractionResults1}). Then, the rectangular image is transformed into a Radon space (Fig.\ref{fig:PLextractionResults2}). The maximum of local maxima is shown by a red star, which is the candidate of the pleural line. (Fig.\ref{fig:PLextractionResults3}) illustrates the result of the extracted pleural line in a rectangular image using inverse Radon transform for the angle and position of the red star. Finally, the image is splitted using the obtained pleural line to define the lung area (Fig.\ref{fig:PLextractionResults4}). Below the yellow line is defined as the lung area and is used for further analysis.

\begin{figure}
    \centering
    \subfloat[]{\includegraphics[scale=0.25]{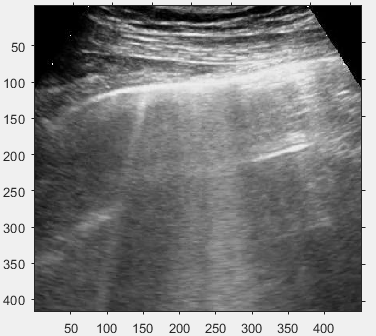}
    \label{fig:fig11a}}
     \subfloat[]{ \includegraphics[scale=0.26]{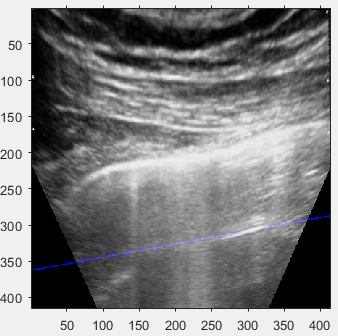}
     \label{fig:fig11b}}
      \subfloat[]{ \includegraphics[scale=0.26]{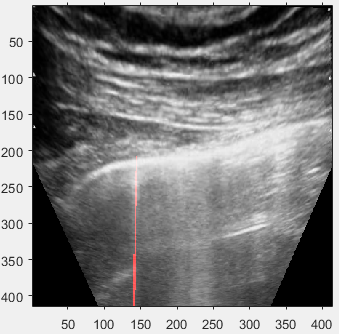}
      \label{fig:fig11c}}
      \subfloat[]{  \includegraphics[scale=0.26]{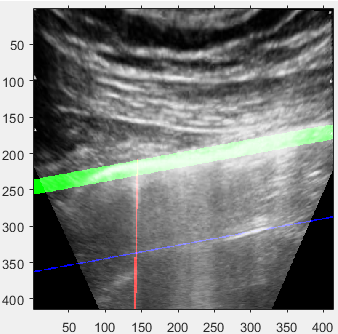}
      \label{fig:fig11d}}
      \subfloat[]{   \includegraphics[scale=0.25]{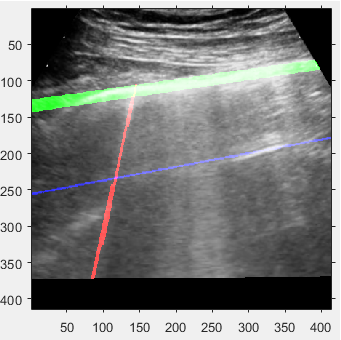}
      \label{fig:fig11e}}
        
    \caption{The results on the suggested method for LUS patterns extraction. (a) Original image, (b) extracted A-lines, (c) extracted B-lines, (d) all patterns projected in the rectangular image, (e) convert the rectangular image back to the trapezoid image }
    \label{fig:fig6-all}
\end{figure}

  The results of A-lines and B-lines extraction are illustrated in Fig  \ref{fig:fig6-all}. Fig. \ref{fig:fig11a} is real sample data in polar coordinates. The A-line extracted in the affined image to Cartesian coordinates is shown in Fig. \ref{fig:fig11b}. An extracted B-line is depicted in Fig. \ref{fig:fig11c}. Fig. \ref{fig:fig11d} shows all extracted patterns together in Cartesian coordinates. Finally, all patterns are highlighted in polar coordinates in  Fig. \ref{fig:fig11e}.

\begin{figure}
    \centering
    {\includegraphics[scale=0.1]{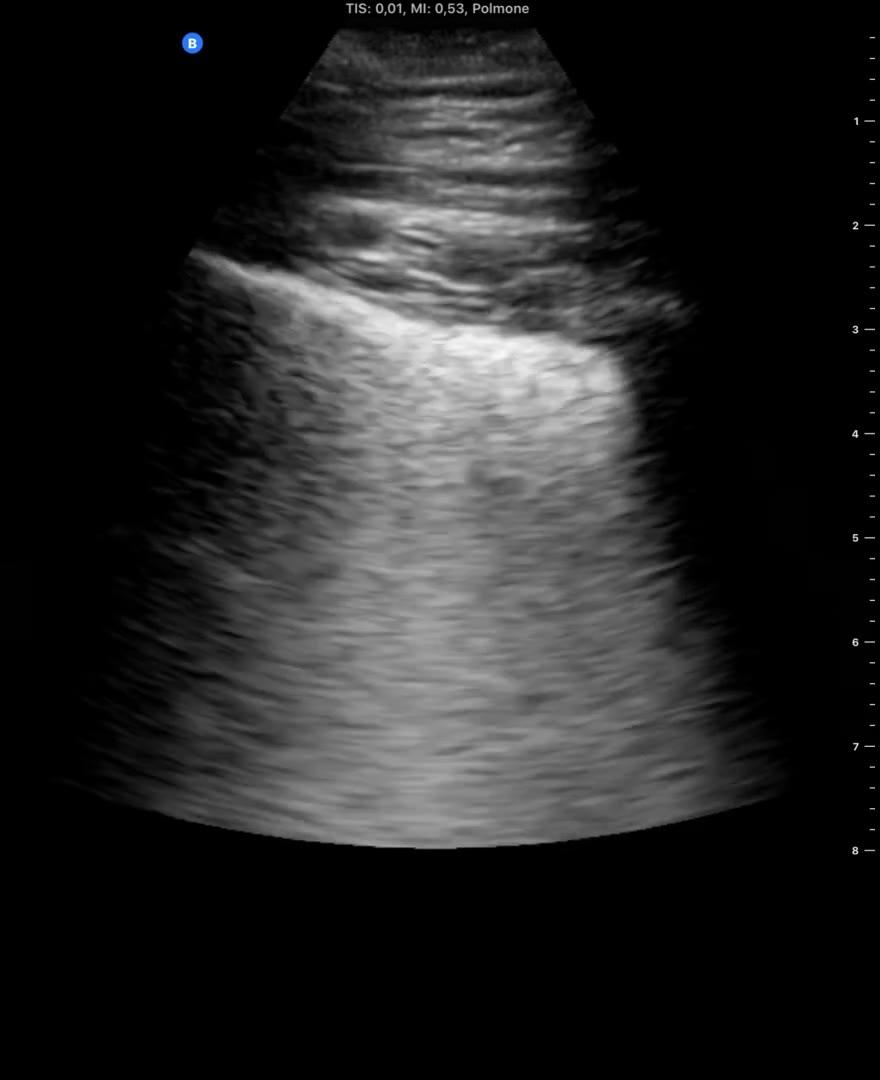}}
     {\includegraphics[scale=0.2]{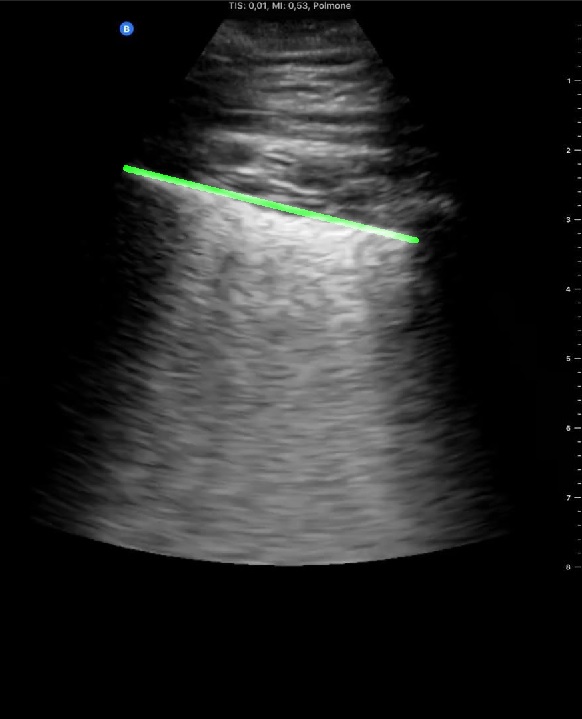}}
      {\includegraphics[scale=0.2]{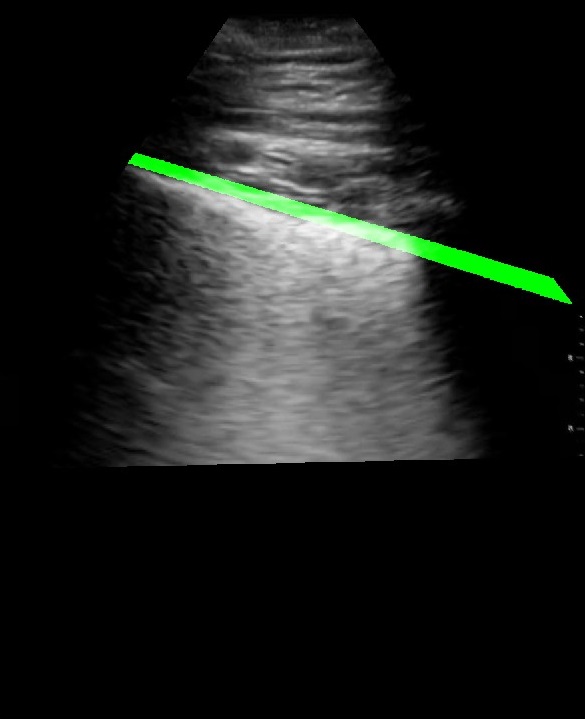}} \\
    {\includegraphics[scale=0.1]{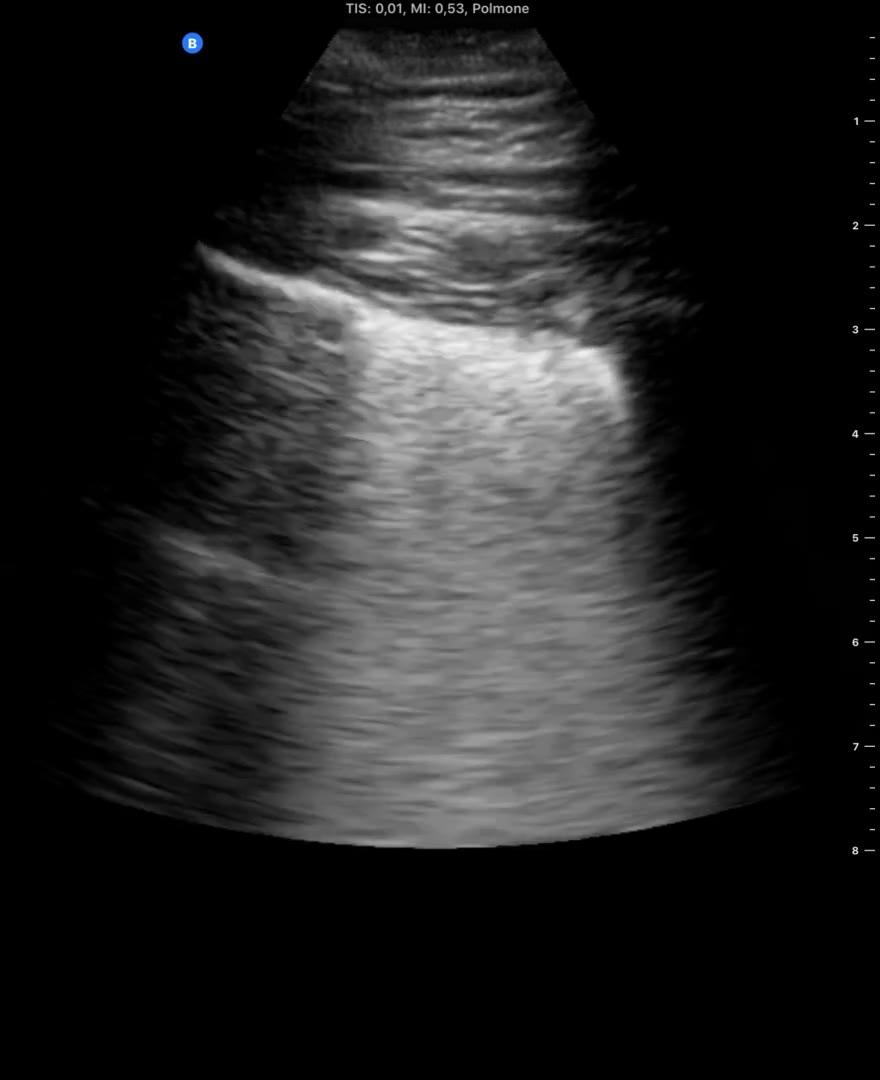}}
 {\includegraphics[scale=0.2]{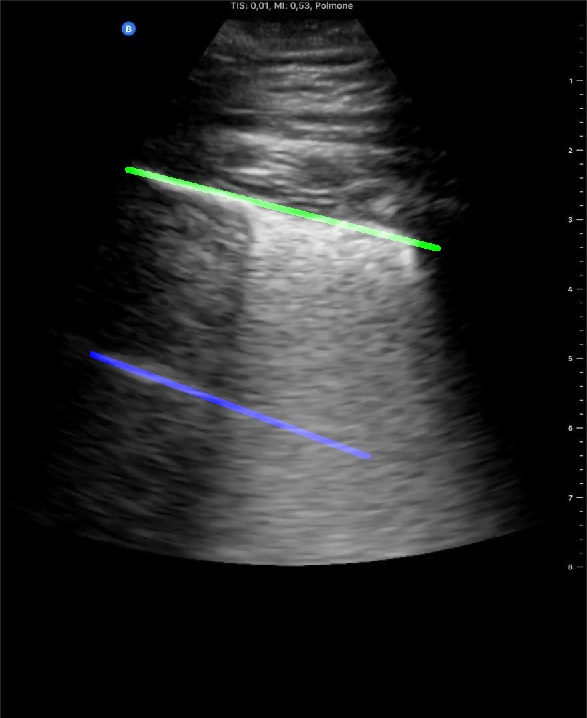}}
    {\includegraphics[scale=0.2]{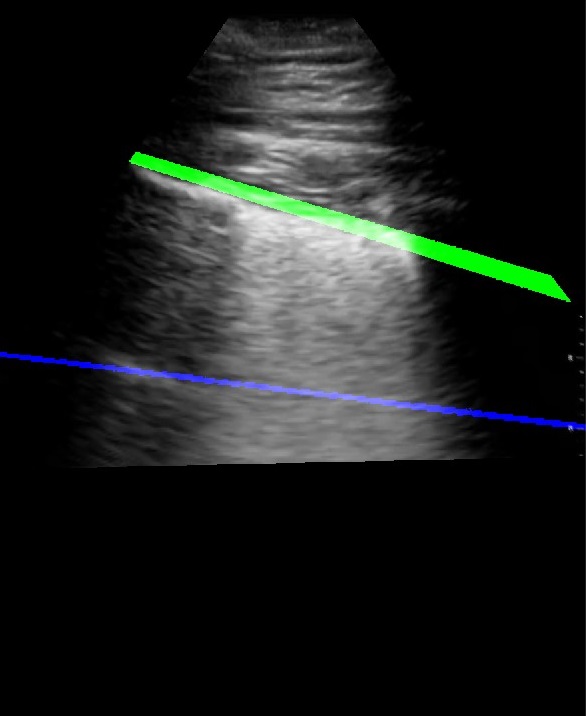}} \\
    {\includegraphics[scale=0.1]{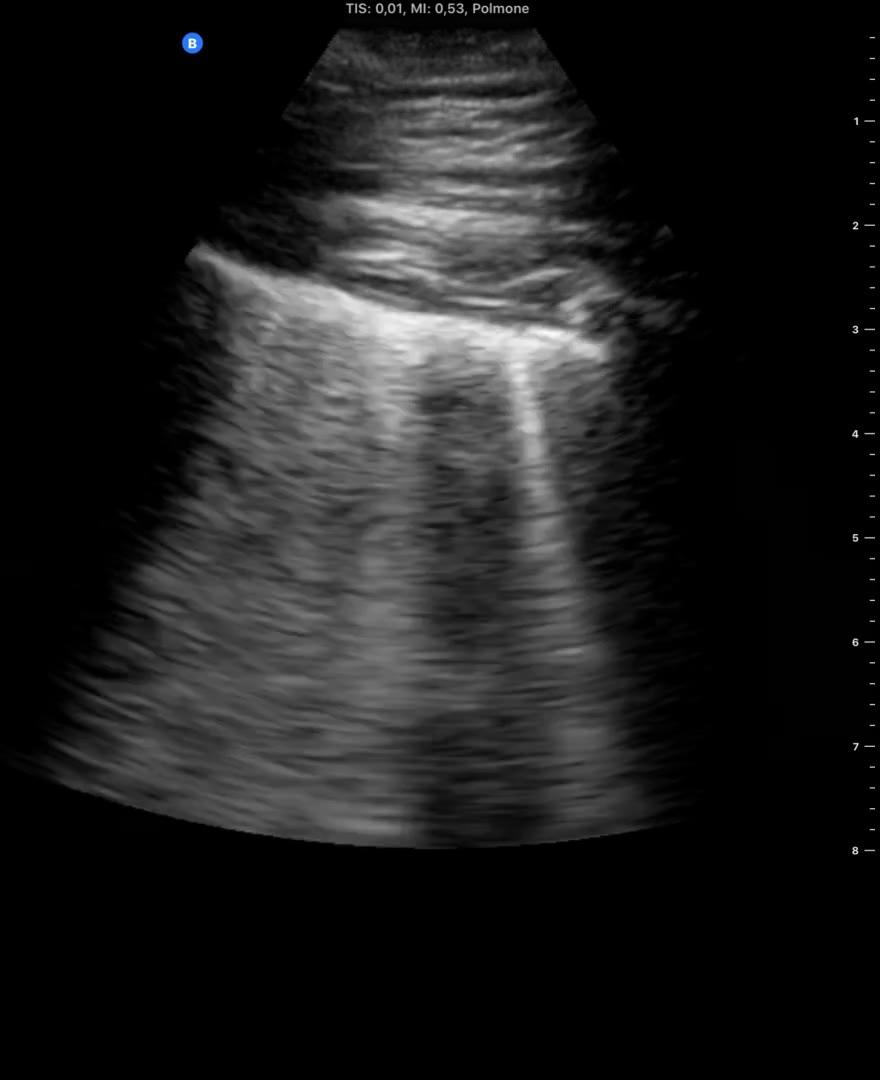}}
    {\includegraphics[scale=0.2]{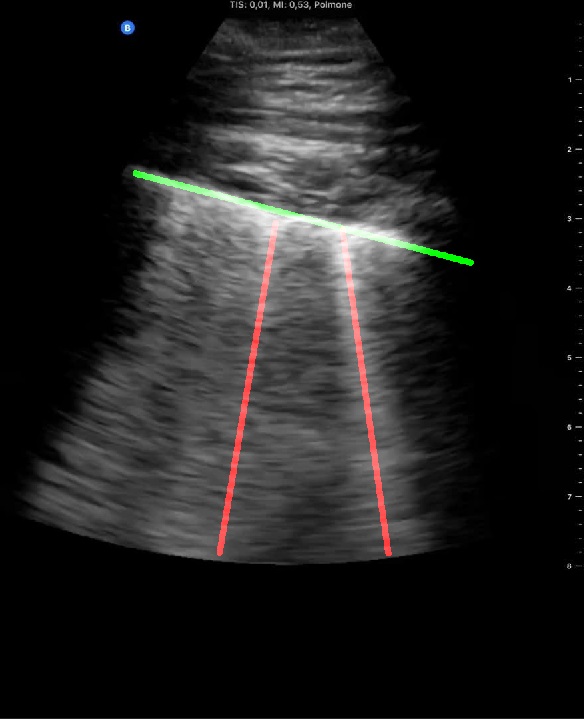}}
    {\includegraphics[scale=0.2]{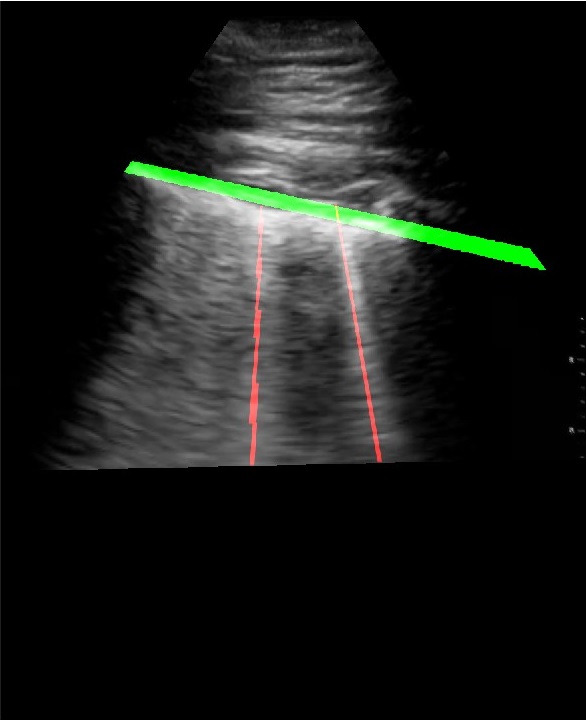}} \\
    {\includegraphics[scale=0.185]{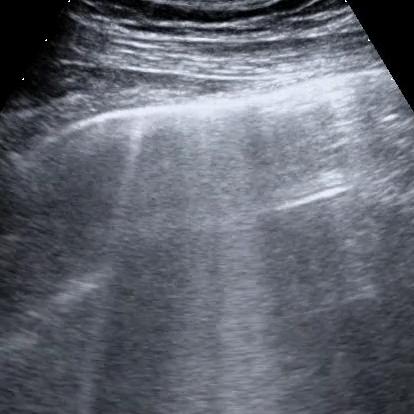}}
    {\includegraphics[scale=0.25]{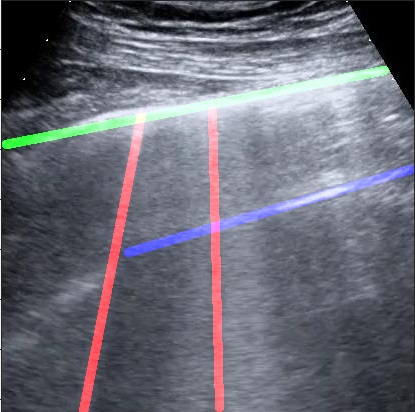}}
    {\includegraphics[scale=0.25]{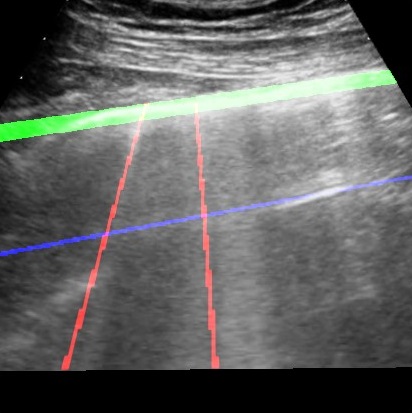}}
    \caption{Detected LUS patterns. First column: original images. Second column: The manually extracted results by an expert. Third column: The obtained results using the proposed algorithm. Pleural lines are in green, A-lines in blue, and B-lines in red.}
    \label{fig:DetectedLUSPatterns}
\end{figure}

Some subjective results are shown in Fig. \ref{fig:DetectedLUSPatterns}. This figure covers images with different patterns inside. It also depicts the comparison of our method with the manually extracted gold standard.  In this figure, the first column illustrates the original images, the second column shows the results manually extracted by an expert, and the third column illustrates the results using the proposed algorithm. Pleural lines are in green, A-lines are in blue, and B-lines are in red.

 The best-obtained F2-score for Pleural line detection is 100\%, 86.29\% for A-lines, and 62.93\% for B-lines extraction. Table \ref{tab: filtersOnRadon} briefly explains the achieved results.
\begin{table}
\centering
\caption{The comparison between obtained results by the proposed method and the work introduced by Anantrasirichai \emph{et al.} \cite{anantrasirichai2017line}. }
    \begin{tabular}{|c|c|c|c|c|}
    \cline{2-5}
    \multicolumn{1}{c|}{} & \multicolumn{3}{c|}{Proposed Method} & Ref.\cite{anantrasirichai2017line}\\
    \cline{2-4}
     \multicolumn{1}{c|}{}& Pleural Line & A-Lines & B-Lines& \\
    \hline
    Precision& 100* & 83.52 & 74.10&- \\
     \hline
    Recall& 100 & 87.02 & 60.65&- \\
     \hline
     F0.5& 100 & 84.19 & 70.95& 54 \\
      \hline
    F1& 100 & 85.23 & 66.70& 40\\
     \hline
     F2& 100 & 86.29 & 62.93& 33\\
         \hline
    \end{tabular}
    
    \label{tab: filtersOnRadon}
    \footnotesize {* all numbers are in percentage.}
\end{table}

\section{Discussion}

In this work, firstly, a Wavelet-based method is utilized to denoise lung ultrasound images. It can be seen that the selection of the threshold value is not sensitive to the selection of the mother wavelet.  Then, the performance of  Radon transform is investigated for pattern detection in LUS images. Patterns are automatically extracted using the local maxima in Radon domain. It is demonstrated that Radon transform is reliable for the extraction of the three most important patterns in LUS images called Pleural line, A-lines, and B-lines. Considering the results of Table \ref{tab: filtersOnRadon}, the proposed algorithm is highly robust in Pleural line detection as its precision and recall are 100 \%.

It is obvious that detecting a higher number of desirable patterns yields a greater recall value. While higher precision means that the number of undesirable detected patterns is low. In our work, it is important not to miss any useful pattern for disease diagnosis. So, higher recall is more important than higher precision. Looking at Table \ref{tab: filtersOnRadon}, our suggested method reached an F2-score of 86.29\%, and 62.93\% for the extraction of A-lines, and B-lines, respectively. For A-lines patterns, F2-Score is higher than F0.5-Score which means that the proposed framework worked well for A-Lines detection. Whereas, F0.5-Score for B-lines detection is higher than F2-Score, indicating that fewer unwanted patterns are detected. Our work achieved satisfactory results for F-Scores compared to previously published work \cite{anantrasirichai2017line}. In future work, we will try to increase the recall value for B-lines detection using more data. We also aim to investigate the performance of deep learning techniques and explore how fast and accurate they are in the topic of this paper.

\section*{Acknowledgment}
The authors would like to thank R. Ketelaars who donated 66 LUS videos recorded from patients suffering from pneumothorax. We also appreciate the contribution of Jannis Born and Lingyi Zhao who helped us with data collection.

\bibliographystyle{unsrtnat}
\bibliography{main}  

\begin{thebibliography}{28}
\providecommand{\natexlab}[1]{#1}
\providecommand{\url}[1]{\texttt{#1}}
\expandafter\ifx\csname urlstyle\endcsname\relax
  \providecommand{\doi}[1]{doi: #1}\else
  \providecommand{\doi}{doi: \begingroup \urlstyle{rm}\Url}\fi

\bibitem[Alvarado and Arce(2016)]{alvarado2016metabolic}
Alcibey Alvarado and Isabel Arce.
\newblock Metabolic functions of the lung, disorders and associated
  pathologies.
\newblock \emph{Journal of clinical medicine research}, 8\penalty0
  (10):\penalty0 689, 2016.

\bibitem[Yoon et~al.(2013)Yoon, Choi, Suh, Jeong, Lee, Park, Kim, and
  Park]{yoon2013tension}
Jeong~Seob Yoon, Si~Young Choi, Jong~Hui Suh, Jin~Yong Jeong, Bae~Young Lee,
  Yong~Gue Park, Chi~Kyung Kim, and Chan~Beom Park.
\newblock Tension pneumothorax, is it a really life-threatening condition?
\newblock \emph{Journal of cardiothoracic surgery}, 8\penalty0 (1):\penalty0
  1--6, 2013.

\bibitem[Zhu et~al.(2020)Zhu, Zhang, Wang, Li, Yang, Song, Zhao, Huang, Shi,
  Lu, et~al.]{zhu2020novel}
Na~Zhu, Dingyu Zhang, Wenling Wang, Xingwang Li, Bo~Yang, Jingdong Song, Xiang
  Zhao, Baoying Huang, Weifeng Shi, Roujian Lu, et~al.
\newblock A novel coronavirus from patients with pneumonia in china, 2019.
\newblock \emph{New England journal of medicine}, 2020.

\bibitem[Yau et~al.(2021)Yau, Gin, Luong, Jue, Abolmaesumi, Tsang, Nair, and
  Tsang]{yau2021point}
Olivia Yau, Ken Gin, Christina Luong, John Jue, Purang Abolmaesumi, Michael
  Tsang, Parvathy Nair, and Teresa~SM Tsang.
\newblock Point-of-care ultrasound in the covid-19 era: A scoping review.
\newblock \emph{Echocardiography}, 38\penalty0 (2):\penalty0 329--342, 2021.

\bibitem[Buonsenso et~al.(2020)Buonsenso, Piano, Raffaelli, Bonadia, Donati,
  and Franceschi]{buonsenso2020novel}
D~Buonsenso, A~Piano, F~Raffaelli, N~Bonadia, K~De~Gaetano Donati, and
  F~Franceschi.
\newblock Novel coronavirus disease-19 pnemoniae: a case report and potential
  applications during covid-19 outbreak.
\newblock \emph{Eur Rev Med Pharmacol Sci}, 24\penalty0 (5):\penalty0 2776--80,
  2020.

\bibitem[Sodhi(2021)]{sodhi2021lung}
Kushaljit~Singh Sodhi.
\newblock Lung mri in children: the road less travelled.
\newblock \emph{Indian Journal of Radiology and Imaging}, 31\penalty0
  (01):\penalty0 237--241, 2021.

\bibitem[Lichtenstein(2014)]{lichtenstein2014lung}
Daniel~A Lichtenstein.
\newblock Lung ultrasound in the critically ill.
\newblock \emph{Annals of intensive care}, 4\penalty0 (1):\penalty0 1--12,
  2014.

\bibitem[Shams et~al.(2021)Shams, Eldesoky, Sobh, Elgaby, and
  Hashem]{shams2021lung}
Doaa Shams, Gehan Eldesoky, Eman Sobh, Soad Elgaby, and Ragia Hashem.
\newblock Lung ultrasound in critically ill patients comparison with bedside
  chest radiography.
\newblock \emph{Azhar International Journal of Pharmaceutical and Medical
  Sciences}, 1\penalty0 (3):\penalty0 118--124, 2021.

\bibitem[Liu et~al.(2019)Liu, Wang, Yang, Lei, Liu, Li, Ni, and
  Wang]{liu2019deep}
Shengfeng Liu, Yi~Wang, Xin Yang, Baiying Lei, Li~Liu, Shawn~Xiang Li, Dong Ni,
  and Tianfu Wang.
\newblock Deep learning in medical ultrasound analysis: a review.
\newblock \emph{Engineering}, 5\penalty0 (2):\penalty0 261--275, 2019.

\bibitem[Soldati et~al.(2020)Soldati, Smargiassi, Inchingolo, Buonsenso,
  Perrone, Briganti, Perlini, Torri, Mariani, Mossolani,
  et~al.]{soldati2020there}
Gino Soldati, Andrea Smargiassi, Riccardo Inchingolo, Danilo Buonsenso, Tiziano
  Perrone, Domenica~Federica Briganti, Stefano Perlini, Elena Torri, Alberto
  Mariani, Elisa~Eleonora Mossolani, et~al.
\newblock Is there a role for lung ultrasound during the covid-19 pandemic?
\newblock \emph{Journal of Ultrasound in Medicine}, 2020.

\bibitem[Smargiassi et~al.(2021)Smargiassi, Soldati, Torri, Mento, Milardi,
  Del~Giacomo, De~Matteis, Burzo, Larici, Pompili, et~al.]{smargiassi2021lung}
Andrea Smargiassi, Gino Soldati, Elena Torri, Federico Mento, Domenico Milardi,
  Paola Del~Giacomo, Giuseppe De~Matteis, Maria~Livia Burzo, Anna~Rita Larici,
  Maurizio Pompili, et~al.
\newblock Lung ultrasound for covid-19 patchy pneumonia: extended or limited
  evaluations?
\newblock \emph{Journal of Ultrasound in Medicine}, 40\penalty0 (3):\penalty0
  521--528, 2021.

\bibitem[Bekgoz et~al.(2019)Bekgoz, Kilicaslan, Bildik, Keles, Demircan,
  Hakoglu, Coskun, and Demir]{bekgoz2019blue}
Burak Bekgoz, Isa Kilicaslan, Fikret Bildik, Ayfer Keles, Ahmet Demircan, Onur
  Hakoglu, Gulhan Coskun, and Huseyin~Avni Demir.
\newblock Blue protocol ultrasonography in emergency department patients
  presenting with acute dyspnea.
\newblock \emph{The American journal of emergency medicine}, 37\penalty0
  (11):\penalty0 2020--2027, 2019.

\bibitem[Bandyk et~al.(2021)Bandyk, Gopireddy, Lall, Balaji, and
  Dolz]{bandyk2021mri}
Mark~G Bandyk, Dheeraj~R Gopireddy, Chandana Lall, KC~Balaji, and Jose Dolz.
\newblock Mri and ct bladder segmentation from classical to deep learning based
  approaches: Current limitations and lessons.
\newblock \emph{Computers in Biology and Medicine}, 134:\penalty0 104472, 2021.

\bibitem[Khan et~al.(2021)Khan, Laghari, and Awan]{khan2021machine}
Abdullah~Ayub Khan, Asif~Ali Laghari, and Shafique~Ahmed Awan.
\newblock Machine learning in computer vision: A review.
\newblock \emph{EAI Endorsed Transactions on Scalable Information Systems},
  8\penalty0 (32):\penalty0 e4--e4, 2021.

\bibitem[Barrientos et~al.(2016)Barrientos, Roman-Gonzalez, Barrientos, Solis,
  Alva, Correa, Pajuelo, Anticona, Lavarello, Casta{\~n}eda,
  et~al.]{barrientos2016filtering}
Franklin Barrientos, Avid Roman-Gonzalez, Ronald Barrientos, Leonardo Solis,
  Alicia Alva, Malena Correa, Monica Pajuelo, Cynthia Anticona, Roberto
  Lavarello, Benjamin Casta{\~n}eda, et~al.
\newblock Filtering of the skin portion on lung ultrasound digital images to
  facilitate automatic diagnostics of pneumonia.
\newblock In \emph{2016 IEEE 36th Central American and Panama Convention
  (CONCAPAN XXXVI)}, pages 1--4. IEEE, 2016.

\bibitem[Karaku{\c{s}} et~al.(2020)Karaku{\c{s}}, Anantrasirichai, Aguersif,
  Silva, Basarab, and Achim]{karakucs2020detection}
Oktay Karaku{\c{s}}, Nantheera Anantrasirichai, Amazigh Aguersif, Stein Silva,
  Adrian Basarab, and Alin Achim.
\newblock Detection of line artifacts in lung ultrasound images of covid-19
  patients via nonconvex regularization.
\newblock \emph{IEEE transactions on ultrasonics, ferroelectrics, and frequency
  control}, 67\penalty0 (11):\penalty0 2218--2229, 2020.

\bibitem[Susanti et~al.(2021)]{susanti2021image}
Hesty Susanti et~al.
\newblock Image processing framework for pleural line (a-line) detection in
  video lung ultrasonography.
\newblock In \emph{2020 IEEE-EMBS Conference on Biomedical Engineering and
  Sciences (IECBES)}, pages 99--102. IEEE, 2021.

\bibitem[Anantrasirichai et~al.(2017)Anantrasirichai, Allinovi, Hayes, Bull,
  and Achim]{anantrasirichai2017line}
Nantheera Anantrasirichai, Marco Allinovi, Wesley Hayes, David Bull, and Alin
  Achim.
\newblock Line detection in speckle images using radon transform and
  $\mathcal{L}$ 1 regularization.
\newblock In \emph{2017 IEEE International Conference on Acoustics, Speech and
  Signal Processing (ICASSP)}, pages 6240--6244. IEEE, 2017.

\bibitem[Li et~al.(2010)Li, Wang, Yu, and Chen]{li2010fetal}
Xiaomin Li, Yuanyuan Wang, Jinhua Yu, and Ping Chen.
\newblock Fetal lung segmentation using texture-based boundary enhancement and
  active contour models.
\newblock In \emph{2010 3rd International Conference on Biomedical Engineering
  and Informatics}, volume~1, pages 264--268. IEEE, 2010.

\bibitem[Zhao et~al.(2022)Zhao, Fong, and Bell]{zhao2022covid}
Lingyi Zhao, Tiffany~Clair Fong, and Muyinatu A~Lediju Bell.
\newblock Covid-19 feature detection with deep neural networks trained on
  simulated lung ultrasound b-mode images.
\newblock In \emph{2022 IEEE International Ultrasonics Symposium (IUS)}, pages
  1--3. IEEE, 2022.

\bibitem[Born et~al.(2021)Born, Wiedemann, Cossio, Buhre, Br{\"a}ndle,
  Leidermann, Goulet, Aujayeb, Moor, Rieck, et~al.]{born2021accelerating}
Jannis Born, Nina Wiedemann, Manuel Cossio, Charlotte Buhre, Gabriel
  Br{\"a}ndle, Konstantin Leidermann, Julie Goulet, Avinash Aujayeb, Michael
  Moor, Bastian Rieck, et~al.
\newblock Accelerating detection of lung pathologies with explainable
  ultrasound image analysis.
\newblock \emph{Applied Sciences}, 11\penalty0 (2):\penalty0 672, 2021.

\bibitem[Ketelaars et~al.(2018)Ketelaars, G{\"u}lpinar, Roes, Kuut, and van
  Geffen]{ketelaars2018ultrasound}
R~Ketelaars, E~G{\"u}lpinar, T~Roes, M~Kuut, and GJ~van Geffen.
\newblock Which ultrasound transducer type is best for diagnosing pneumothorax?
\newblock \emph{Critical ultrasound journal}, 10\penalty0 (1):\penalty0 1--9,
  2018.

\bibitem[Piscaglia et~al.(2020)Piscaglia, Stefanini, Cantisani, Sidhu, Barr,
  Berzigotti, Chammas, Correas, Dietrich, Feinstein,
  et~al.]{piscaglia2020benefits}
Fabio Piscaglia, Federico Stefanini, Vito Cantisani, Paul~S Sidhu, Richard
  Barr, Annalisa Berzigotti, Maria~Cristina Chammas, Jean-Michel Correas,
  Christoph~Frank Dietrich, Steven Feinstein, et~al.
\newblock Benefits, open questions and challenges of the use of ultrasound in
  the covid-19 pandemic era. the views of a panel of worldwide international
  experts.
\newblock \emph{Ultraschall in der Medizin-European Journal of Ultrasound},
  41\penalty0 (03):\penalty0 228--236, 2020.

\bibitem[Gupta et~al.(2018)Gupta, Taneja, and Chand]{gupta2018performance}
Manoj Gupta, Heena Taneja, and Laxmi Chand.
\newblock Performance enhancement and analysis of filters in ultrasound image
  denoising.
\newblock \emph{Procedia computer science}, 132:\penalty0 643--652, 2018.

\bibitem[Georgieva et~al.(2021)Georgieva, Petrov, and
  Zlatareva]{georgieva2021medical}
Veska Georgieva, Plamen Petrov, and Dora Zlatareva.
\newblock Medical image processing based on multidimensional wavelet
  transforms-advantages and trends.
\newblock In \emph{AIP Conference Proceedings}, volume 2333, page 020001. AIP
  Publishing LLC, 2021.

\bibitem[Georgieva and Grozeva(2020)]{georgieva2020comparative}
Veska Georgieva and Anna Grozeva.
\newblock A comparative analysis of various filters for noise reduction in
  cardiac ultrasound images.
\newblock In \emph{2020 XI National Conference with International Participation
  (ELECTRONICA)}, pages 1--4. IEEE, 2020.

\bibitem[Farahi et~al.(2022)Farahi, Casals, Sarrafzadeh, Zamani, Ahmadi,
  Behbood, and Habibian]{farahi2022beat}
Maria Farahi, Al{\'\i}cia Casals, Omid Sarrafzadeh, Yasaman Zamani, Hooran
  Ahmadi, Naeimeh Behbood, and Hessam Habibian.
\newblock Beat-to-beat fetal heart rate analysis using portable medical device
  and wavelet transformation technique.
\newblock \emph{Heliyon}, page e12655, 2022.

\bibitem[Jiang(2020)]{jiang2020statistical}
Wenxin Jiang.
\newblock Statistical formulas for f measures.
\newblock \emph{arXiv preprint arXiv:2012.14894}, 2020.

\end{thebibliography}
\end{document}